\documentclass[aps,prl,twocolumn,showpacs]{revtex4}
\usepackage{amssymb,amsbsy,graphicx,times,subfigure,psfig,float}
\usepackage[latin1]{inputenc}
\newcommand{\gr}[1]{\boldsymbol{#1}}
\newcommand{\be}{\begin{equation}}
\newcommand{\ee}{\end{equation}}
\newcommand{\bea}{\begin{eqnarray}}
\newcommand{\eea}{\end{eqnarray}}

\newcommand{\N}{{\cal N}}
\newcommand{\D}{\Delta}
\newcommand{\abs}[1]{\left\vert#1\right\vert}
\renewcommand{\det}{{\rm Det}\,}

\begin{document}
%%%%%%%%%%%%%%%%%%%%%%%%%%%%%%%%%%%%%%%%%%%%%%%%%%%
\title{Determination of continuous variable entanglement by purity measurements}
%%%%%%%%%%%%%%%%%%%%%%%%%%%%%%%%%%%%%%%%%%%%%%%%%%%
\author{Gerardo Adesso}
\author{Alessio Serafini}
\author{Fabrizio Illuminati}
\affiliation{Dipartimento di Fisica ``E. R. Caianiello'',
Universit\`a di Salerno, INFM UdR di Salerno, INFN Sezione di Napoli,
Gruppo Collegato di Salerno,
Via S. Allende, 84081 Baronissi (SA), Italy}
\date{February 25, 2004}
%%%%%%%%%%%%%%%%%%%%%%%%%%%%%%%%%%%%%%%%%%%%%%%%
\begin{abstract}
We classify the entanglement of two--mode Gaussian states according to
their degree of total and partial mixedness. We derive exact bounds
that determine maximally and minimally entangled states
for fixed global and marginal purities. This characterization
allows for an experimentally reliable estimate
of continuous variable entanglement based on measurements of purity.
\end{abstract}
%%%%%%%%%%%%%%%%%%%%%%%%%%%%%%%%%%%%%%%%%%%%%%%%
%%%%%%%%%%%%%%%%%%%%%%%%%%%%%%%%%%%%%%%%%%%%%%%%
\maketitle
%%%%%%%%%%%%%%%%%%%%%%%%%%%%%%%%%%%%%%%%%%%%%%%%%%%

%%%%%%%%%%%%%%%%%%%%%%%%%%%%%%%%%%%%%%%%%%%%%%%%%
Quantum entanglement of Gaussian states constitutes a fundamental
resource in continuous variable (CV) quantum information
\cite{Book}. Therefore, the quest for a theoretically satisfying
and experimentally realizable quantification of the entanglement
for such states stands as a major issue in the field. On the
theoretical ground, a proper, computable quantitative
characterization of the entanglement of Gaussian states is
provided by the logarithmic negativity \cite{werner}. Experimental
schemes to determine the entanglement of Gaussian states have been
proposed both in the two--mode \cite{kim} and in the multipartite
\cite{vloock} instance. However, these schemes are based on
homodyne detections, and require a full reconstruction of all the
second moments of the Gaussian field. \par In this work, we
present a theoretical framework to estimate the entanglement of
two--mode Gaussian states by the knowledge of the total and of the
two partial purities. This is achieved by deriving analytical {\it
a priori} upper and lower bounds on the logarithmic negativity for
fixed values of the global and marginal purities. We then show
that the set of entangled Gaussian states is tightly contained
between two extremal surfaces of maximally and minimally entangled
states. This quantification allows for a simple strategy to
measure the entanglement of Gaussian states with reliable
experimental accuracy. In fact, measurements of global and
marginal purities do not require the demanding reconstruction of
the full covariance matrix and can be performed directly by
exploiting the technology of quantum networks \cite{network}.\par
Let us consider a two--mode continuous variable system, described
by the Hilbert space ${\cal H}={\cal H}_{1}\otimes{\cal H}_{2}$
resulting from the tensor product of the Fock spaces ${\cal
H}_{k}$'s. We denote by $a_{k}$ the annihilation operator acting
on the space ${\cal H}_{k}$. Likewise, $\hat
x_{k}=(a_{k}+a^{\dag}_{k})/\sqrt{2}$ and $\hat
p_{k}=-i(a_{k}-a^{\dag}_{k})/\sqrt{2}$ are the quadrature phase
operators of the mode $k$, the corresponding phase space variables
being $x_{k}$ and $p_{k}$.\par In the following, we will make use
of the Wigner quasi--probability representation $W(x_{i},p_{i})$,
defined as the Fourier transform of the symmetrically ordered
characteristic function. In Wigner phase space picture, the tensor
product ${\cal H}={\cal H}_{1}\otimes{\cal H}_{2}$  results in the
direct sum $\Gamma=\Gamma_{1}\oplus\Gamma_{2}$ of the related
phase spaces $\Gamma_{i}$'s. A symplectic transformation acting on
the global phase space $\Gamma$ corresponds to a unitary operator
acting on $\cal H$ \cite{simon}. We will refer to a transformation
$S_{l} = S_{1} \oplus S_{2}$, with each $S_{i} \in Sp_{(2,\mathbb
R)}$ acting on $\Gamma_{i}$, as to a ``local symplectic
operation'', corresponding to a ``local unitary transformation''
$U_{l}= U_{1}\otimes U_{2}$. The set of Gaussian states is defined
as the set of states with Gaussian Wigner function
\begin{equation}
W(X)=\frac{\,{\rm e}^{-\frac{1}{2}X\boldsymbol{\sigma}^{-1}X^{T}}}{\pi
\sqrt{{\rm Det}\,\boldsymbol{\sigma}}}{\:,}
\label{wigner}
\end{equation}
where $X\equiv (x_{1},p_{1},x_{2},p_{2})\in\Gamma$,
and we will denote by $\hat{X}$ the vector of
operators $(\hat{x}_{1},\hat{p}_{1},\hat{x}_{2},\hat{p}_{2})$.
First moments have been neglected, since they
can be set to zero by means of a local unitary transformation.
Second moments form the covariance matrix $\boldsymbol{\sigma}$
of the Gaussian state
$
\sigma_{ij}\equiv\frac{1}{2}\langle \hat{X}_i \hat{X}_j +
\hat{X}_j \hat{X}_i \rangle -
\langle \hat{X}_i \rangle \langle \hat{X}_j \rangle
$.
For simplicity, in what follows
$\boldsymbol{\sigma}$ will refer both to the Gaussian state
and to its covariance matrix.
It is convenient to express
$\boldsymbol{\sigma}$ in terms of the three $2\times 2$
matrices $\boldsymbol{\alpha}$, $\boldsymbol{\beta}$,
$\boldsymbol{\gamma}$
\begin{equation}
\boldsymbol{\sigma}\equiv\left(\begin{array}{cc}
\boldsymbol{\alpha}&\boldsymbol{\gamma}\\
\boldsymbol{\gamma}^{T}&\boldsymbol{\beta}
\end{array}\right)\, .
\label{espre}
\end{equation}\par
\noindent Heisenberg uncertainty principle can be expressed as \cite{simon}
\begin{equation}
\boldsymbol{\sigma}+\frac{i}{2}\boldsymbol{\Omega}\ge 0 \; ,
\label{bonfide}
\end{equation}
where $\boldsymbol{\Omega}\equiv\gr{\omega}\oplus\gr{\omega}$
is the usual symplectic form with $\omega_{ij}=\delta_{i j-1}-\delta_{i j+1}$,
$i,j=1,2$.
Ineq.~(\ref{bonfide}) can be recast as a
constraint on the $Sp_{(4,{\mathbb R})}$
invariants $\Delta \equiv \,{\rm Det}\,\gr{\alpha}+
\,{\rm Det}\,\gr{\beta}+2\,{\rm Det}\,\gr{\gamma}$,
and ${\rm Det}\,\gr{\sigma}$ \cite{simon}
\be
\Delta\le\frac{1}{4}
+ 4\,{\rm Det}\,\boldsymbol{\sigma}
\, . \label{sympheis}
\ee\par
In general, the Wigner function transforms as a scalar
under symplectic operations,
while the covariance matrix $\boldsymbol{\sigma}$ transforms
according to
$\boldsymbol{\sigma}\rightarrow
S^{T}\boldsymbol{\sigma}S$, with $S \in Sp_{(4,\mathbb R)}$.
As it is well known \cite{duan}, for any covariance
matrix $\boldsymbol{\sigma}$ there exists a local
canonical operation $S_{l}=S_{1}\oplus S_{2}$ that
recasts $\boldsymbol{\sigma}$ in the ``standard form''
$\boldsymbol{\sigma}_{sf}$ with $\gr{\alpha}=\,{\rm diag}\,\{a,a\}$,
$\gr{\beta}=\,{\rm diag}\,\{b,b\}$,
$\gr{\gamma}=\,{\rm diag}\,\{c_{+},c_{-}\}$,
where $a$, $b$, $c_{+}$, $c_{-}$ are determined
by the four local symplectic
invariants ${\rm Det} \,
\boldsymbol{\sigma}$,
${\rm Det}\,\boldsymbol{\alpha}$, ${\rm Det}
\,\boldsymbol{\beta}$, and
${\rm Det}\,\boldsymbol{\gamma}$. \par
Any bipartite Gaussian state $\gr{\sigma}$ can always be written as
$\gr{\sigma}=S^T \gr{\nu} S $ for some
$S\in Sp_{4,\mathbb{R}}$ and $\gr{\nu}=
\,{\rm diag}\,\{n_{-},n_{-},n_{+},n_{+}\}$. The quantities
$n_{\mp}$ constitute the {\em symplectic spectrum} of $\gr{\sigma}$;
they are determined by the global symplectic invariants \cite{holevo,sirkaz}
\be
2n_{\mp}^2=\Delta\mp\sqrt{\Delta^2-4\,{\rm Det}\,\gr{\sigma}} \, .
\label{eigen}
\ee
In terms of $n_{\mp}$ Ineq.~(\ref{sympheis}) becomes simply $n_{-}\ge1/2$.
\par
We will characterize the mixedness of a quantum state $\varrho$ by its
purity $\mu\equiv\,{\rm Tr}\,\varrho^2$. For a $n$--mode
Gaussian state $\gr{\sigma}$
the purity is simply evaluated integrating the Wigner function, yielding
$\mu=1/(2^n\,\sqrt{{\rm Det} \, \gr{\sigma}})$. \par
As for the entanglement, we recall that the positivity of the partially
transposed (PPT) state $\tilde{\gr{\sigma}}$
is equivalent to separability for any two--mode Gaussian state $\gr{\sigma}$
\cite{simon}.
In terms of
symplectic invariants, partial transposition corresponds to
flipping the sign of ${\rm Det}\,\gr{\gamma}$, so that $\Delta$ turns into
$\tilde{\Delta}=\Delta-4\,{\rm Det}\,\gr{\gamma}$.
The symplectic spectrum $\tilde{n}_{\mp}$ of $\tilde{\gr{\sigma}}$ is simply found
inserting $\tilde{\Delta}$ for $\Delta$ in Eq.~(\ref{eigen}).
If $\tilde{n}_{-}$ is the smallest symplectic eigenvalue of the partially transposed
covariance matrix $\tilde{\gr{\sigma}}$, a state $\gr{\sigma}$ is separable
if and only if
\begin{equation}
\tilde{n}_{-}\ge1/2 \label{sep} \; .
\label{lowest}
\end{equation}
A {\it bona fide} measure of entanglement
for two--mode Gaussian states should thus be a monotonically
decreasing function of $\tilde{n}_{-}$,
quantifying the violation of inequality (\ref{lowest}). A
computable entanglement monotone for generic two-mode
Gaussian states is provided by
the logarithmic negativity $E_{\N}=\max\{0,-\ln(2\tilde{n}_{-})\}$
\cite{werner}.
For symmetric Gaussian states, i.e. states whose standard form
is characterized by $\boldsymbol{\alpha} = \boldsymbol{\beta}$, another
computable entanglement monotone is provided by the entanglement
of formation \cite{giedke}. However, in this subcase the two measures provide
the same characterization of entanglement and are fully equivalent.
Therefore, from now on we will adopt the logarithmic negativity
to quantify the entanglement of two-mode Gaussian states. \par
We now show that a generic state in standard form can be reparametrized
in terms of the $Sp_{(4,\mathbb R)}$
invariants $\mu$ (the global purity) and $\Delta$,
and of the $Sp_{(2,\mathbb R)}\oplus Sp_{(2,\mathbb R)}$
invariants $\mu_1$ and $\mu_2$,
where $\mu_{i}$ denotes the
purity of the reduced state in mode $i$ ($i=1,2$).
For a generic two-mode Gaussian state $\gr{\sigma}_{sf}$
we thus have
\begin{eqnarray}
  \mu_1 &=& \frac{1}{2a}\,, \quad \mu_2 \,\,=\,\, \frac{1}{2b}\,, \label{mu12}\\
  \frac{1}{16\mu^2} &=& \det{\gr{\sigma}} = (a b)^2-a b (c_+^2+c_-^2)+(c_+ c_-)^2\,, \label{dets}\\
  \Delta &=& a^2+b^2+2c_+ c_-\,. \label{deltas}
\end{eqnarray}
Eqs.~(\ref{mu12}-\ref{deltas}) are easily inverted to provide the
following parametrization
\begin{eqnarray}
  a &=& \frac{1}{2\mu_1}\,, \quad b \,\,=\,\, \frac{1}{2\mu_2}\,, \label{gab} \\
c_{\pm}&=&\frac12 \sqrt{\mu_1 \mu_2 \left[\left(\D-\frac{(\mu_1 - \mu_2)^2}
{4 \mu_1^2 \mu_2^2}\right)^2-\frac{1}{4\mu^2}\right]} \; \pm \, \epsilon \nonumber \\
\label{gc}
\end{eqnarray}
$${\rm with}\quad
\epsilon \, \equiv \,
\frac18 \sqrt{\frac{\left[(\mu_1+\mu_2)^2 - 4 \mu_1^2 \mu_2^2 \D\right]^2}{\mu_1^3 \mu_2^3}
-\frac{4\mu_1 \mu_2}{\mu^2}}\, .
$$
The global and marginal purities range from $0$ to $1$,
constrained by the condition
\begin{equation}\label{ineqmumui}
\mu \ge \mu_1 \mu_2 \; ,
\end{equation}
a direct consequence of Heisenberg uncertainty relations. It
implies that no Gaussian LPTP (less pure than product) states
exist, at variance with the case of two--qubit systems
\cite{adesso}.
Eqs.~(\ref{eigen},\ref{gab}, \ref{gc}) determine the smallest symplectic
eigenvalue of the covariance matrix $\gr{\sigma}$ and of its
partial transpose $\tilde{\gr{\sigma}}$
\begin{equation}
  2n_{-}^2 = \Delta-\sqrt{\Delta^2
-{\frac{1}{4 \mu^2}}}\,, \quad
  2\tilde{n}_{-}^2 = \tilde{\Delta}-\sqrt{\tilde{\Delta}^2
-{\frac{1}{4 \mu^2}}}\,, \label{n1}
\end{equation}
where $\tilde{\Delta} = -\D + 1/2\mu_1^2 + 1/2\mu_2^2$. This
parametrization describes physical states if the radicals in
Eqs.~(\ref{gc}, \ref{n1}) exist and Ineq.~(\ref{sympheis}),
expressing the Heisenberg principle, is satisfied. All these
conditions can be combined and recast as upper and lower bounds on
the invariant $\D$
\begin{eqnarray}
  & & \frac{1}{2 \mu} + \frac{(\mu_1 - \mu_2)^2}{4 \mu_1^2 \mu_2^2}
  \,\, \le\,\, \D \nonumber\\
  &\le& \min \left\{ \frac{(\mu_1 + \mu_2)^2}{4 \mu_1^2 \mu_2^2}
  - \frac{1}{2 \mu} \; , \; \frac14 \left(1+\frac{1}{\mu^2}\right)  \right\}
  \, . \label{deltabnd}
\end{eqnarray}
The invariant $\D$ has a direct physical
interpretation: at given global and marginal
purities, it determines the amount of entanglement
of the state. In fact, one has
\begin{equation}
  \left. \frac{\partial\ \tilde{n}^2_{-}}{\partial\ \D}
  \right|_{\mu_1,\,\mu_2,\,\mu} \,
 = \; \frac12 \left(
\frac{\tilde{\Delta}}{\sqrt{\tilde{\Delta}^2
-{\frac{1}{4 \mu^2}}}} -1 \right) \,
> 0 \, .
\end{equation}
The smallest symplectic eigenvalue of the partially transposed state
is strictly monotone in $\D$. Therefore the entanglement of a generic
Gaussian state $\gr{\sigma}$ with global purity $\mu$ and marginal
purities $\mu_{1,2}$ strictly increases with decreasing $\D$.
Since $\D$ has both lower and upper
bounds, due to Ineq.~(\ref{deltabnd}),
not only maximally but also \emph{minimally}
entangled Gaussian states exist. This is an important
result concerning the relation between entanglement and purity of
quantum states: the entanglement of a Gaussian state is tightly
bound by the amount of global and marginal purities, with only
one remaining degree of freedom related to the invariant $\D$.\par
We now aim to characterize extremal (maximally or minimally)
entangled Gaussian states for fixed global and marginal purities.
Let us first consider the states
saturating the lower bound in Eq.~(\ref{deltabnd}),
which entails \emph{maximal} entanglement.
They are Gaussian maximally entangled mixed states (GMEMS), admitting the
following parametrization
\begin{equation} \label{gnsm}
a=\frac{1}{2 \mu_{1}}, \; b=\frac{1}{2 \mu_{2}}, \; c_{\pm}= \pm \frac12
\sqrt{\frac{1}{\mu_1 \mu_2}-\frac{1}{\mu}} \; .
\end{equation}
We now recall that Gaussian squeezed thermal states are states
of the form $\gr{\sigma}=S_{r}\gr{\nu}S_{r}$, where
$S_{r}$ is the symplectic representation of the two--mode squeezing operator
${\rm Exp}\,[r(a_{1}a_{2}-a_{1}^{\dag}a_{2}^{\dag})/2]$, while
$\gr{\nu}=\,{\rm diag}\,\{n_{-},n_{-},n_{+},n_{+}\}$. These states
are in standard form with $a=n_{-}\cosh^{2}r+n_{+}\sinh^{2}r$,
$b=n_{+}\cosh^{2}r+n_{-}\sinh^{2}r$, $\;c_{+}=-c_{-}=
(n_{-}+n_{+})\sinh2r/2$. In the pure case ($n_{\mp}=1/2$) they
reduce to two--mode squeezed vacua. We thus find
that states of the form of Eq.~(\ref{gnsm}) are non--symmetrical
squeezed thermal states with
$\tanh2r=(\mu_1\mu_2-\mu_1^2\mu_2^2/\mu)^{1/2}/(\mu_1+\mu_2)$.
These states are \emph{separable} in the range
\begin{equation}\label{gnsmsep}
\mu \le \frac{\mu_1 \mu_2}{\mu_1 + \mu_2 - \mu_1 \mu_2}\,.
\end{equation}
In such a {\it separable region} in the space of purities,
no entanglement can occur for states of the form of Eq.~(\ref{gnsm}),
while, outside this region, they are GMEMS.
We now consider the class of states that saturate the upper
bound in Eq.~(\ref{deltabnd}). They determine the class
of Gaussian least entangled mixed states (GLEMS).
Violation of Ineq.~(\ref{gnsmsep}) implies that
$\left( 1 + 1/\mu^2 \right)/4 \le
(\mu_1 + \mu_2)^{2}/4 \mu_1^{2} \mu_2^{2} - 1/2\mu$. Therefore,
outside the separable region, GLEMS fulfill
\begin{equation}\label{glm}
\D = \frac14 \left(1+\frac{1}{\mu^2}\right) \,.
\end{equation}
Eq.~(\ref{glm}) expresses saturation of Heisenberg
relation (\ref{sympheis}). We thus find that the
most semiclassical states of minimum quantum uncertainty
are Gaussian least entangled states.
GLEMS in standard form are characterized by
\begin{eqnarray}
c_{\pm} \!\!&=&\!\! \frac18 \sqrt{\mu_1 \mu_2
\left[-\frac{4}{\mu^2}+\left(1+\frac{1}{\mu^2}-\frac{(\mu_1-\mu_2)^2}{\mu_1^2
\mu_2^2}\right)^2\right]}\nonumber \\
\!\!&\pm&\!\!
\frac{1}{8\mu}\sqrt{-4\mu_1\mu_2+\frac{\left[\left(
1+\mu^2\right)\mu_1^2\mu_2^2-\mu^2{\left(\mu_1+\mu_2\right)}^2\right]
^2}{\mu^2\mu_1^3\mu_2^3}} \; . \nonumber \\
\end{eqnarray}
According to the PPT criterion, GLEMS are
separable only for $\mu \le \mu_1 \mu_2 / \sqrt{\mu_1^2 + \mu_2^2
- \mu_1^2 \mu_2^2}$, so that in the range
\begin{equation}\label{gnslsep}
\frac{\mu_1 \mu_2}{\mu_1 + \mu_2 - \mu_1 \mu_2} < \mu \le
\frac{\mu_1 \mu_2}{\sqrt{\mu_1^2 + \mu_2^2 - \mu_1^2 \mu_2^2}}
\end{equation}
both separable and entangled states can be found.
The very narrow region defined by
Ineq.~(\ref{gnslsep}) is the only {\em coexistence region}
between entangled and separable Gaussian mixed states.
Furthermore, Ineq.~(\ref{deltabnd})
leads to the following constraint on the purities
\begin{equation}\label{upineqmumui}
\mu \le \frac{\mu_1 \mu_2}{\mu_1 \mu_2 + \abs{\mu_1-\mu_2}}\,.
\end{equation}
For purities which saturate Ineq.~(\ref{upineqmumui}), GMEMS and
GLEMS coincide and we have a unique class of states depending only
on the marginal purities $\mu_{1,2}$. They are Gaussian
maximally entangled states for fixed marginals (GMEMMS).
The maximal entanglement of
a Gaussian state decreases rapidly with increasing
difference of marginal purities, in analogy with
finite-dimensional systems \cite{adesso}. For symmetric states $(\mu_1=\mu_2)$
Ineq.~(\ref{upineqmumui}) reduces to the trivial bound $\mu \le 1$
and GMEMMS reduce to pure two--mode squeezed states.
\begin{figure}[t!]
  \includegraphics[width=6cm]{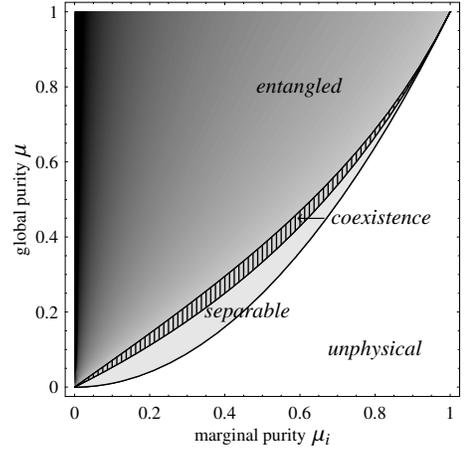}\\
  \caption{Summary of entanglement properties of symmetric Gaussian
  states with given global and marginal purities.
  In the entangled region, the average logarithmic
  negativity Eq.~(\ref{average}) is depicted, growing from gray to black.
  The dashed area is the coexistence region of separable and entangled states.}
  \label{fig2D}
\end{figure}
We can summarize the previous results in the following scheme, classifying
all the two-mode Gaussian physical
states according to their degree of global and marginal purities
\begin{eqnarray}
  \mu_1 \mu_2 \; \le \; \mu \; \le \; \frac{\mu_1 \mu_2}{\mu_1 + \mu_2 - \mu_1 \mu_2}
  \!\!&\Rightarrow&\!\! \textrm{separable} \nonumber\\
  \frac{\mu_1 \mu_2}{\mu_1 + \mu_2 - \mu_1 \mu_2} < \mu \le
  \frac{\mu_1 \mu_2}{\sqrt{\mu_1^2 + \mu_2^2 - \mu_1^2 \mu_2^2}}
\!\!&\Rightarrow&\!\! \textrm{coexistence} \nonumber\\
  \frac{\mu_1 \mu_2}{\sqrt{\mu_1^2 + \mu_2^2 - \mu_1^2 \mu_2^2}} <
  \mu \le \frac{\mu_1 \mu_2}{\mu_1 \mu_2 + \abs{\mu_1-\mu_2}}
  \!\!&\Rightarrow&\!\! \textrm{entangled} \nonumber\\
  \label{entsum}
\end{eqnarray}
Knowledge of the global and marginal purities thus
accurately characterizes the entanglement of Gaussian states,
providing strong sufficient conditions and analytical bounds.
As we will now show, marginal and global purities allow
also an accurate quantification of entanglement.
Outside the separable region, GMEMS attain maximum
logarithmic negativity $E_{{\N}max}$
\bea
E_{{\N}max}(\mu_{1,2},\mu) = -\frac12\log \Bigg[-\frac{1}{\mu}
  + \left(\frac{\mu_1+\mu_2}{2\mu_1^2 \mu_2^2} \right)
  \nonumber \\
  \times  \left(\mu_1+\mu_2 -
  \sqrt{(\mu_1+\mu_2)^2-\frac{4 \mu_1^2 \mu_2^2}{\mu}} \right) \Bigg] \, ,
\label{enmax}
\eea
while, in the entangled region (see Eq.~(\ref{entsum})), GLEMS
acquire minimum logarithmic negativity $E_{{\N}min}$
\bea
E_{{\N}min}(\mu_{1,2},\mu) = - \frac12 \log \Bigg[\frac{1}{\mu_1^2}+\frac{1}{\mu_2^2}-\frac{1}{2\mu^2} -
  \frac12  \nonumber\\
- \sqrt{\left( \frac{1}{\mu_1^2}+\frac{1}{\mu_2^2}-
\frac{1}{2\mu^2} - \frac12 \right)^2 - \frac{1}{\mu^2}} \; \Bigg] \, .
\label{enmin}
\eea

\begin{figure}[t!]
  \includegraphics[width=6.5cm]{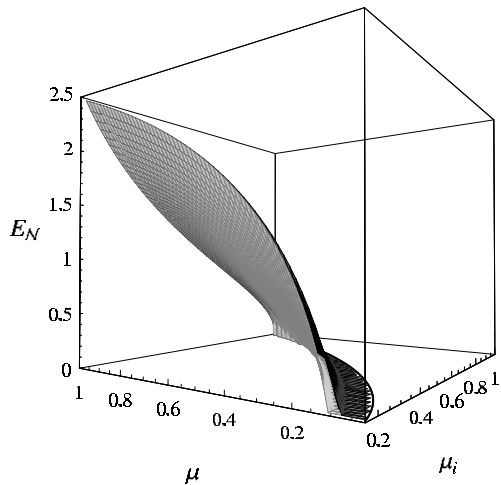}\\
  \caption{Upper and lower bounds on the logarithmic
  negativity as functions of the global and marginal purities of
  symmetric Gaussian states. The black (gray) surface represents GMEMS
  (GLEMS).}
  \label{fig3D}
\end{figure}

Knowledge of $\D$ (i.e. of the full covariance matrix) would allow
for an exact quantification of the entanglement. However, we will
now show that an estimate based only on the knowledge of the
experimentally measurable global and marginal purities turns out
to be quite accurate. We can in fact quantify the entanglement of
Gaussian states with given global and marginal purities by the
``average logarithmic negativity'' $\bar{E}_{\N}$ defined as \be
\bar{E}_{\N} (\mu_{1,2},\mu) \equiv
\frac{E_{{\N}max}(\mu_{1,2},\mu)+E_{{\N}min}(\mu_{1,2},\mu)}{2} \;
. \label{average} \ee We can then also define the relative error
$\delta \bar{E}_{\N}$ on $\bar{E}_{\N}$ as
\begin{equation}\label{deltaen}
\delta \bar{E}_{\N} (\mu_{1,2},\mu) \equiv
\frac{E_{Nmax}(\mu_{1,2},\mu)
-E_{Nmin}(\mu_{1,2},\mu)}{E_{Nmax}(\mu_{1,2},\mu)
+E_{Nmin}(\mu_{1,2},\mu)}\,.
\end{equation}
It is easily seen that this error decreases both with increasing
global purity and decreasing marginal purities, i.e. with
increasing entanglement. For ease of graphical display, let us
consider the important case of symmetric Gaussian states, for
which the reduction $\mu_1 = \mu_2 \equiv \mu_i$ occurs.
\begin{figure}[th]
  \includegraphics[width=6.5cm]{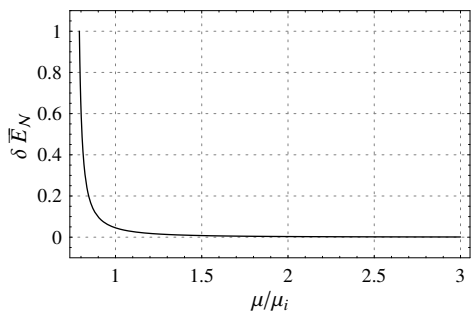}\\
  \caption{The relative error $\delta \bar{E}_{\N}$
Eq.~(\ref{deltaen}) on the average logarithmic negativity as a
function of the ratio $\mu / \mu_i$, plotted at $\mu=0.5$.}
  \label{figerror}
\end{figure}

Fig.~\ref{fig2D} shows the classification of the entanglement of
symmetric states depending on their global and marginal purities.
Notice in particular the very narrow region of coexistence of
separable and entangled states. In Fig.~\ref{fig3D},
$E_{{\N}min}(\mu_i,\,\mu)$ of Eq.~(\ref{enmin}) and
$E_{{\N}max}(\mu_i,\,\mu)$ of Eq.~(\ref{enmax}) are plotted versus
$\mu_i$ and $\mu$. In the case $\mu=1$ the upper and lower bounds
correctly coincide, since for pure states the entanglement is
completely quantified by the marginal purity. For mixed states
this is not the case, but, as the plot shows, knowledge of the
global and marginal purities strictly bounds the entanglement both
from above and from below.

The relative error $\delta \bar{E}_{\N} (\mu_i,\,\mu)$ given by
Eq.~(\ref{deltaen}) is plotted in Fig.~\ref{figerror} as a
function of the ratio $\mu / \mu_i$. It decays exponentially, and
falls below $5\%$ for $\mu > \mu_i$. Thus detection of genuinely
entangled states is always assured by this method, except at most
for a small set of states with very weak entanglement (states with
$E_{\N} \lesssim 1$). Moreover, the accuracy is even greater in
the general non-symmetric case $\mu_1 \neq \mu_2$, because the
maximal entanglement decreases in such an instance. The above
analysis demonstrates that the average logarithmic negativity
$\bar{E}_{\N}$ is a reliable estimate of the logarithmic
negativity $E_{\N}$, improving as the entanglement increases. This
allows for an accurate quantification of CV entanglement by
knowledge of the global and marginal purities. The latter
quantities may be in turn amenable to direct experimental
determination by exploiting the technology of quantum networks
\cite{network}, even without homodyning\cite{cerf}. The present
work thus may provide a powerful operative characterization and
quantification of the entanglement of generic Gaussian states.
\smallskip

\noindent Financial support from INFM, INFN, and MIUR under
national project PRIN-COFIN 2002 is acknowledged.

\end{document}